\documentclass[aps,reprint,nofootinbib, amsmath,amssymb,showpacs,showkeys,superscriptaddress]{revtex4-1}

\usepackage{graphicx}  
\usepackage{subfig} 
\usepackage{natbib}  
\usepackage{amsmath}  
\usepackage[colorlinks=true,linkcolor=blue,citecolor=blue]{hyperref}  
\usepackage[top=1.5in, bottom=1.5in, left=1in, right=1in]{geometry}
\usepackage{caption}
\usepackage{amsfonts}
\usepackage{xcolor}
\usepackage{booktabs}
\usepackage{multirow}
\usepackage{bbm}
\usepackage{soul,url}
\usepackage{afterpage}

\begin{document} 

\title{A novel stacked hybrid autoencoder for imputing LISA data gaps}

\author{Ruiting Mao} \affiliation{Department of Statistics,
  University of Auckland, Auckland 1010, New Zealand}

\author{Jeong Eun Lee} \affiliation{Department of Statistics,
  University of Auckland, Auckland 1010, New Zealand}

\author{Matthew C. Edwards} \affiliation{Department of Statistics,
  University of Auckland, Auckland 1010, New Zealand} 
  

\begin{abstract}
The Laser Interferometer Space Antenna (LISA) data stream will contain gaps with missing or unusable data due to antenna repointing, orbital corrections, instrument malfunctions, and unknown random processes.  We introduce a new deep learning model to impute data gaps in the LISA data stream.  The stacked hybrid autoencoder combines a denoising convolutional autoencoder (DCAE) with a bi-directional gated recurrent unit (BiGRU).  The DCAE is used to extract relevant features in the corrupted data, while the BiGRU captures the temporal dynamics of the gravitational-wave signals.  We show for a massive black hole binary signal, corrupted by data gaps of various numbers and duration, that we yield an overlap of greater than 99.97\% when the gaps do not occur in the merging phase and greater than 99\% when the gaps do occur in the merging phase.  However, if data gaps occur during merger time, we show that we get biased astrophysical parameter estimates, highlighting the need for ``protected periods'', where antenna repointing does not occur during the predicted merger time.  
    
\end{abstract}


\maketitle

\section{Introduction}

The Laser Interferometer Space Antenna (LISA) \cite{lisa:2017, baker2019laser} is a space-borne gravitational-wave observatory under development by the European Space Agency (ESA) and the National Aeronautics and Space Administration (NASA) \cite{bender1998pre}.  It could observe a wealth of possible gravitational-wave (GW) sources in the range 0.1 mHz to 1 Hz \cite{colpi2024lisadefinitionstudyreport}, such as those from massive black hole binaries (MBHBs) \cite{berti2006gravitational, hughes2002untangling, sesana2005gravitational, vecchio2004lisa}, extreme-mass ratio inspirals (EMRIs) \cite{Gair_2017,Babak_2017}, and galactic binaries (GBs) \cite{Cornish_2017,Willems_2008,Littenberg_2019}. A fundamental distinction between LISA and the ground-based detectors (Advanced LIGO, Advanced Virgo, and KAGRA) lies in the fact that LISA will be capable of detecting sources that will persist within its frequency range for extended durations, spanning from hours to years. 

The observatory will be operational for a nominal duration of four years, with potential extensions. However, due to factors such as antenna repointing, orbital corrections, instrument malfunctions, and other stochastic processes, the LISA data stream is expected to contain missing or unusable data \cite{Amaro_Seoane_2021}. To mitigate the impact of these data disturbances, it may be necessary to exclude the affected segments, which will result in gaps within the usable data streams. These data gaps will induce nonstationarity in the underlying noise process of the LISA instrument, consequently, the covariance matrix commonly employed in our statistical models will cease to be diagonal \cite{wang2024windowinpaintingdealingdata}. Therefore, the estimation of astrophysical parameters will be biased unless these data gaps are adequately addressed. Furthermore, the discrete Fourier transform (DFT) of data containing gaps is susceptible to spectral leakage, thereby affecting both the gravitational wave signal and the stochastic noise \cite{carré2010effect,PhysRevD.102.084062}. This degradation becomes increasingly significant as the frequency of the source decreases \cite{Baghi_2019}.

The operational data from LISA Pathfinder (LPF), launched in December 2015, underscores the necessity to consider the data gaps. It shows that there may indeed be disruptions in the data stream. Simple patterns for gaps in LISA were discussed in \cite{carré2010effect}, which conducted a large-scale Monte Carlo parameter estimation simulation for GBs. Studies on data gaps in LISA have highlighted its impact on detectability and parameter estimation \cite{Baghi_2019}. Dey et al.~\cite{Dey_2021} investigated the two types of gaps in MBHBs, showing that unscheduled gaps have a greater impact compared to scheduled ones. 

To address this challenge, classic techniques from other domains for handling missing data, such as linear interpolation or mean imputation, often prove inadequate when dealing with the intricacies of GW signals \cite{Blackman_2014}. The phase-coherent signals from ultra-compact binaries (UCBs) can be used as calibration sources to measure the duration of data gaps in LISA independently. However, these benchmarks are heavily dependent on the realization of the galactic population of the UCB \cite{PhysRevD.98.043008}. The application of apodization, i.e., implementing a window function to the signal before taking the Discrete Fourier Transform (DFT), is widely used in signal processing and spectral analysis to mitigate the effects of spectral leakage~\cite{carré2010effect, Dey_2021,wang2024windowinpaintingdealingdata}. Indeed, treating the remaining data segments as independent measurements may lead to modeling errors, and there is no ideal smoothing parameter for the window function. A Bayesian augmentation method was applied to treat missing data as auxiliary variables and samples them along with the parameters of interest, providing a statistically consistent way to handle gaps in GBs while improving sampling efficiency and mitigating spectral leakage effects \cite{Baghi_2019}. However, the high reliance on the parametric model assumption on time series will be challenging when extending it to other LISA sources, such as MBHBs. Longer duration gaps were considered in \cite{10.1093/mnras/stab3314,Blelly_2020}, where a nonparametric inpainting algorithm grounded in sparse representation was implemented to attenuate the effects of gaps on the galactic binary signal in the frequency domain. This model-independent approach is heavily reliant on noise modeling, resulting in potential inaccuracies with the intricate patterns and dependencies inherent in gravitational wave data \cite{zhao2023dawningneweragravitational}.

Deep learning has made considerable strides in processing data from GW observations, achieving significant success in this area. Various types of neural networks have been applied to detect and characterize GW signals, such as the convolutional neural network (CNN) in \cite{PhysRevD.101.104003,Krastev_2021,george2018deep,PhysRevD.103.024025}, conditional variational autoencoders (VAEs) in \cite{Gabbard2019BayesianPE}, the transformer-based extraction network in \cite{Zhao_2023}, and also the generative adversarial network (GAN) to generate simulated GW signals \cite{Jadhav_2023, eccleston_2024, lopez:2022, powell:2023}.  Furthermore, deep generative models make it possible to accelerate the generation of GW waveforms \cite{chua:2019, katz:2021, Liao_2021}. Likelihood-free methods can be applied to approximate posteriors through deep learning techniques as well, such as VAE \cite{Gabbard_2018} and normalizing flows \cite{green:2020,Dax_2021,Langendorff_2023}. Deep learning techniques and algorithms have also been applied to various other fascinating challenges in gravitational wave research, like glitch classification, glitch cancelation, and gravitational wave bursts \cite{AI_2023}.

Data gaps can be considered as noisy data within the data stream, making it reasonable to develop an autoencoder (AE) to address missing data in the data stream~\cite{wang2024deep} as well as to manage high-dimensional signals with a small training set \cite{PhysRevD.109.083002}. Denoising autoencoders (DAE) are extensively utilized in analyzing GW observed data to identify and eliminate noise, thereby producing a denoised or reconstructed signal for subsequent analysis \cite{bacon2022denoising,PhysRevD.108.043024,morawski2021anomaly}. Given the time series nature of data streams from GW detectors, recurrent neural networks (RNNs) are frequently integrated with AE to perform detection and denoising tasks. Based on a convolutional autoencoder (CAE), long-short-term memory networks (LSTMs) \cite{hochreiter1997long} were incorporated as layers in both the encoder and decoder of AEs, demonstrating superior performance in reconstructing samples affected by various anomalies \cite{moreno2021sourceagnosticgravitationalwavedetectionrecurrent}. Identical methodologies are evident in \cite{Shen_2019}, yielding promising outcomes in signal extraction against highly noisy backgrounds. A comparable structure is illustrated in \cite{raikman2024gwak}, where the input of the autoencoder is concentrated by the output of LSTMs and the signal itself. The asymmetric addition of RNN layers in AE is also observed in \cite{chatterjee2021extraction,Xu:2024jbo}, where LSTM layers are incorporated into the decoder component. 
Nevertheless, the substantial memory requirements and training latency associated with long sequence GW signals necessitate the segmentation of data streams into smaller fragments when incorporating RNN layers within AE, which poses a potential risk of losing information pertaining to the overall structure of signals.

The prevalence of missing data in time series analysis is a common issue across various disciplines, and encoder-decoder RNNs techniques have achieved substantial success in the reconstruction of time series data. By adding LSTM layers in the encoder of DAE, higher imputation performance in multivariate time series can be seen compared with multi-directional recurrent neural networks \cite{8982996}. A spatio-temporal LSTM convolutional autoencoder method was proposed to fill gaps in satellite retrieval \cite{9884482}; Jia et al.\cite{8217773} proposed a stacked autoencoder-based imputation method that employs two loss functions: one exclusive for the internal LSTM layers and the other for the overall DAE with the same bottleneck layer. Similar to LSTMs, gated recurrent units, also known as GRUs, are utilized extensively in the process of gap imputation to avoid vanishing or exploding gradient issues \cite{cho2014learning,gupta2017instability,che2016recurrentneuralnetworksmultivariate,9374359}. Moreover, in \cite{9221727}, empirical results demonstrated that GRU exhibits a $29.29\%$ increase in processing speed compared to LSTM when applied to an identical dataset, and it also yields superior performance with smaller training data. Alonso et al. \cite{alonso2024gap} conducted empirical investigations with different unidirectional (simple RNN, GRU, LSTM) and bi-directional (BiSRNN, BiGRU, BiLSTM) RNN layers in DAE, and showed that BiGRU layers have the best performance with a low reconstruction error in industry data. A comprehensive comparative analysis of various hybrid models incorporating DAEs and RNN layers for short-term market forecasting has been conducted in \cite{abu2024comparative}, which indicates that BiGRUs or GRUs generally exhibit superior performance relative to other hybrid models. DAEs incorporating GRU layers are prevalent in numerous imputation tasks \cite{CHEN2021120451,IKHLASSE202211565,s23249697,10498920}; these models typically integrate GRU layers within the DAE and utilize a single loss function for imputing sequences significantly shorter than GW signals.

The concept of stacked denoising autoencoders was introduced in \cite{vincent2010stacked}, demonstrating that a locally applied unsupervised criterion yields a more effective representation of the preceding layer. Sequence to sequence AE and GRU based hybrid model were developed in \cite{Rai2021ARA}, which gave a more robust result in short-term solar power forecasting. Drawing inspiration from these pioneering methodologies, we propose an innovative stacked hybrid autoencoder architecture featuring a locally trained denoising convolutional autoencoder (DCAE) as the encoder, complemented by bi-directional gated recurrent unit (BiGRU) layers in the decoder. This model employs two loss functions to effectively execute imputation tasks. This framework is particularly advantageous for GW signal analysis within a LISA data stream compromised by intermittent disruptions. We will introduce our proposed model, the Bi-directional Gated Recurrent Unit Convolutional Autoencoder (BiGRU-CAE), in Section \ref{sec:model_stru} following the methodology background in Section \ref{sec:methodology}. Then we demonstrate our method using a simple toy example in Section \ref{sec:toy_model} and finally show its general applicability on an MBHB source within the LISA framework in Section \ref{sec:massive_black_holes_application}. We give some concluding remarks and future directions in Section \ref{sec:discussion}.

In summary, the novelty of this work is in several directions: 
\begin{itemize}
 \item We propose a novel imputation method for long-time series that is scalable enough for GW data analysis. It stacks a DCAE and BiGRU. In the LISA data stream, signals tend to be longer and can not be cut into pieces if we want to consider gaps at some specific periods, such as at merger time for MBHB signals. Instead of using the bottleneck layer of the denoising autoencoder as the input of the BiGRU, we apply two stacked hybrid models, where a DCAE is trained in the encoder to make the computation stable and efficient.
 \item Our model can tackle both scheduled gaps and unscheduled gaps. Long-duration unscheduled gaps have also been considered in our analysis. Instead of using frequent daily random gaps (not longer than 1 hour), we investigate the impact of long-duration unscheduled gaps, which last 6 hours on average, and apply deep learning imputation to fill these gaps. 
 \item It is a common practice to employ overlap and Signal-to-Noise Ratio (SNR) loss as metrics for assessing performance in denoising applications. We further elaborate on the results of parameter estimation to demonstrate the precision of the parameter estimation in the reconstructed signal. 
 \item Addressing gaps that occur during the merger time of MBHB signals presents a significant challenge. 
To the best of the author's knowledge, this is the first time this problem has been considered. We illustrate the discrepancies in the parameter estimation of the reconstructed signal when such gaps occur during versus outside the merger time.
\end{itemize}

\section{Methodology}\label{sec:methodology}
\subsection{Denoising convolutional autoencoder}

Autoencoders~\cite{autoencoderidea} are a class of neural network architectures that aim to learn efficient representations (encodings) of input data, typically for the purpose of dimension reduction or feature learning. Unlike other types of neural network, the target of training is the input itself. This self-supervised learning approach allows autoencoders to be trained without annotated data, which can be particularly advantageous in scenarios where labeled data are scarce or expensive to obtain.

The fundamental architecture of an autoencoder consists of two main parts: an encoder and a decoder. The encoder compresses the input into a lower-dimensional latent space (also called the bottleneck layer), capturing the most salient features of the data. The decoder then reconstructs the input data from this compressed representation, aiming to minimize the difference between the original input and its reconstruction. The performance of an autoencoder is typically measured by the reconstruction error (usually mean squared error), which quantifies how well the decoder can reconstruct the input from the reduced encoding.

\begin{figure}[h!]
    \centering
    \includegraphics[width=0.5\textwidth]{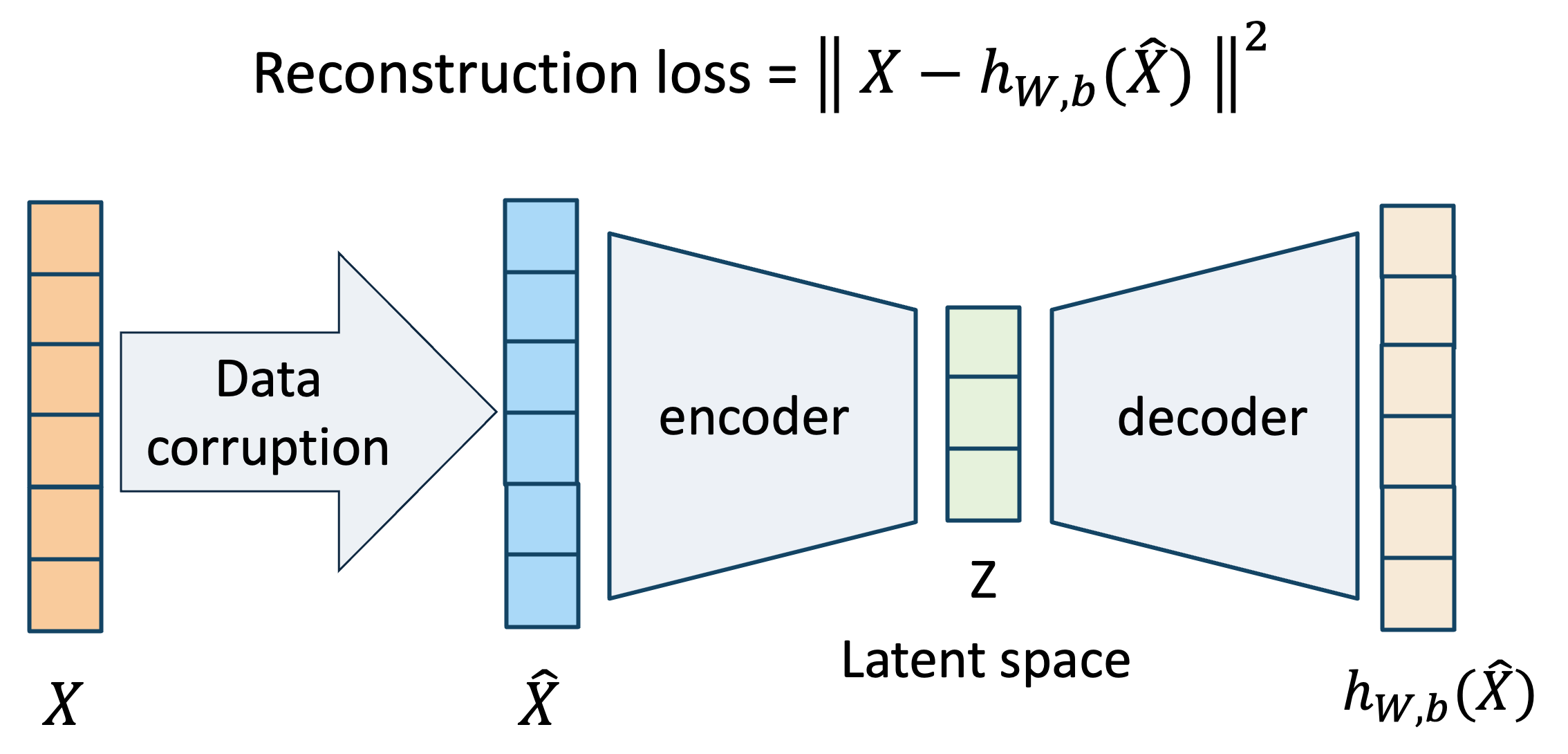}
    \caption{Denoising Autoencoders (DAE) structure. $h_{W,b}$ represents the trained model, which has learned to reconstruct the original input $X$ from its corrupted version $\hat{X}$. }
    \label{fig:DAE}
\end{figure}

Particularly notable among autoencoder variants are denoising autoencoders (DAEs), seen in Fig.~\ref{fig:DAE}, designed to improve data quality by correcting input corrupted with some form of noise. This approach not only helps reduce the noise, but also forces the autoencoder to learn more robust and essential features of the data \cite{8616075}, and prevents the model from learning the identity function. DAEs achieve this by first intentionally corrupting clean input data $X$ to $\hat{X}$, training the model to recover the original uncorrupted input \cite{10.1145/1390156.1390294}. Convolutional layers are computationally efficient in handling large-scale inputs and allow neural networks to learn features that are spatially invariant. Therefore, denoising convolutional autoencoders (DCAEs) are proposed \cite{JMLR:v11:vincent10a}, adding convolutional layers have shown better performance than nonconvolutional neural networks in image processing \cite{DCAEP}. 

\subsection{Gated recurrent unit}

As part of our network architecture, we make use of the gated recurrent unit (GRU) \cite{cho:2014,chung2014empiricalevaluationgatedrecurrent}.  The GRU is a type of recurrent neural network (RNN) that uses a gating mechanism to decide what information passes to the output, thus filtering out irrelevant information.  Standard RNNs suffer from the vanishing gradient problem, where the gradients of the loss function become close to zero and are backpropagated through the neural network \cite{basodi:2020}.  GRUs can avoid this issue altogether using update and reset gates to regulate the flow of information, allowing them to learn long-term time dependencies. GRUs are faster than LSTMs in low-complexity sequences \cite{Cahuantzi_2023}.

\begin{figure}[h!]
    \centering
    \includegraphics[width=0.35\textwidth]{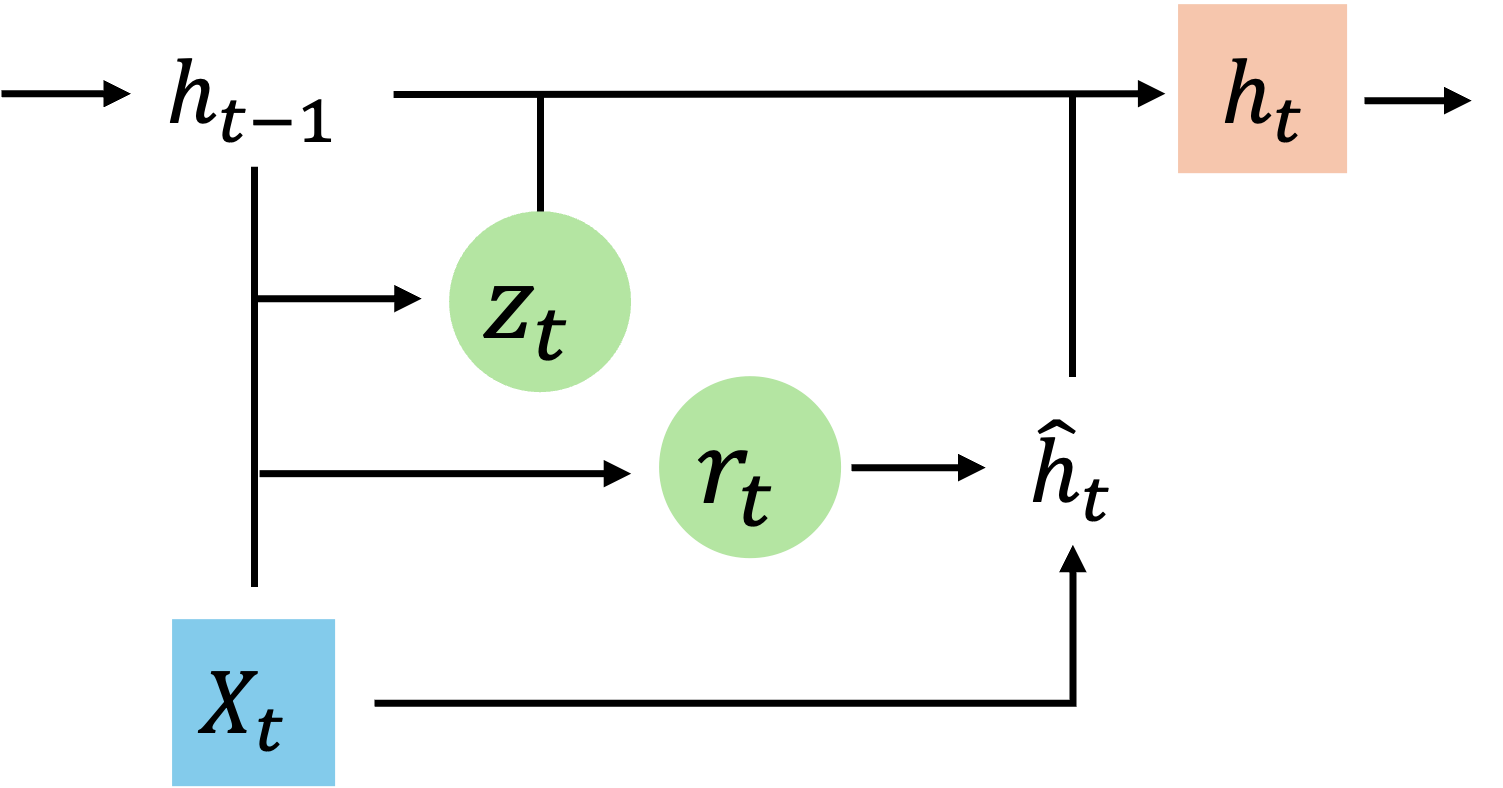}
    \caption{Gated Recurrent Unit (GRU) structure. $r_t$ represents the reset gate, $z_t$ is the update gate, $h_{t-1}$ is the past hidden state, $X_t$ is the observation, $\hat{h}_t$ is the candidate hidden state and $h_t$ is the hidden state. }
    \label{fig:GRU}
\end{figure}

As can be seen in Fig.~\ref{fig:GRU}, the reset gate determines how much information from the past hidden state should be forgotten.  It outputs a value between 0 and 1, where 0 means forget everything and 1 means remember everything.  The output is used to determine a candidate hidden state, or the new information. The update gate determines how much weight to put on the past hidden state and how much weight to put on the candidate hidden state.  It outputs a number between 0 and 1, where 0 means put all the weight on the previous hidden state and 1 means put all of the weight on the candidate hidden state. The standard GRU $GRU(x_t,h_{t-1})$ is formulated:
\begin{widetext}
\begin{align}
r_t &= \text{sigmoid}(W_r x_t + U_r h_{t-1} + b_r) \tag{Reset gate} \\
z_t &= \text{sigmoid}(W_z x_t + U_z h_{t-1} + b_z) \tag{Update gate} \\
\hat{h}_t &= \text{tanh}(W_h x_t + U_h(r_t \odot h_{t-1}) + b_h) \tag{Candidate hidden state} \\
h_t &= (1 - z_t) \odot h_{t-1} + z_t \odot \hat{h}_t \tag{Updated hidden state}
\end{align}
\end{widetext}
The trainable parameters for this neural network are $(W_z, U_z, b_z, W_r, U_r, b_r, W_h, U_h, b_h)$. In addition to the standard GRU, the bi-directional GRU (BiGRU) enhances the model's ability to capture context from both past and future states. In a BiGRU, two separate GRU layers are used: one processes the input sequence in a forward direction, while the other processes it in reverse. 
\begin{align} \label{eq:back_forward_GRU}
\overrightarrow{h}_t &= GRU_f(x_t,\overrightarrow{h}_{t-1}) \tag{Forward}\\
\overleftarrow{h}_t &= GRU_b(x_t,\overleftarrow{h}_{t+1}) \tag{Backward}\\
h_t^{bi} &= [\overrightarrow{h}_t; \overleftarrow{h}_t] \tag{Combined BiGRU output}
\end{align}
This structure allows the network to utilize information from both directions, improving performance on tasks that benefit from understanding the time dependence surrounding each input.

\section{Bi-directional Gated Recurrent Unit Convolutional Autoencoder (BiGRU-CAE)}
Our BiGRU-CAE hierarchical neural network contains a denoising convolutional autoencoder (DCAE) and bi-directional gated recurrent unit (BiGRU) layers to deal with the complicated and long-duration LISA data. The denoising autoencoder model leverages the representation learning capabilities of CNNs to extract relevant features for the whole structure from the input data, while the BiGRU component captures the temporal dynamics of the GW signals. By training this end-to-end hybrid model to reconstruct clean GW signals from noisy detector data, it is able to effectively denoise and extract the underlying GW signals, simplifying the following signal processing procedure. Similar hybrid deep learning architectures have been shown to demonstrate superior performance compared to other techniques to recover GW signals from noisy detector data, such as CNN-LSTM in LIGO-Virgo data analysis \cite{Chatterjee_2021} and DENSE-LSTM model in Taiji data analysis \cite{Xu:2024jbo}.

\subsection{Model structure}\label{sec:model_stru}
The proposed hybrid model has two components. The DCAE component consists of a three-layer 1D convolutional encoder, followed by a fully-connected layer and a three-layer 1D transposed convolutional decoder, simplifying the computation and reducing the input signal's high dimension. To further enhance the autoencoder's efficiency, we employ larger strides in the convolutional layers, thereby eliminating the need for pooling operations \cite{7780459} and the complexity of learning the overall structure of the sequence. This design choice enables the autoencoder to acquire a resilient representation of the input data, essential for further processing. The DCAE functions as the encoder component of the entire hybrid model, with its output regarded as a refined bottleneck.

In the decoder section of the model, bi-directional Gated Recurrent Unit (BiGRU) layers are considered. The purpose of these layers is to infer and capture the temporal patterns embedded within the sequential segments of the signal, thereby refining the partially denoised output generated by the DCAE. The BiGRU part consists of two BiGRU layers, followed by a fully connected layer, enabling the model to effectively learn and retain the temporal dependencies inherent in the input sequence. 

\begin{figure*}
    \centering
    \includegraphics[width=1\textwidth]{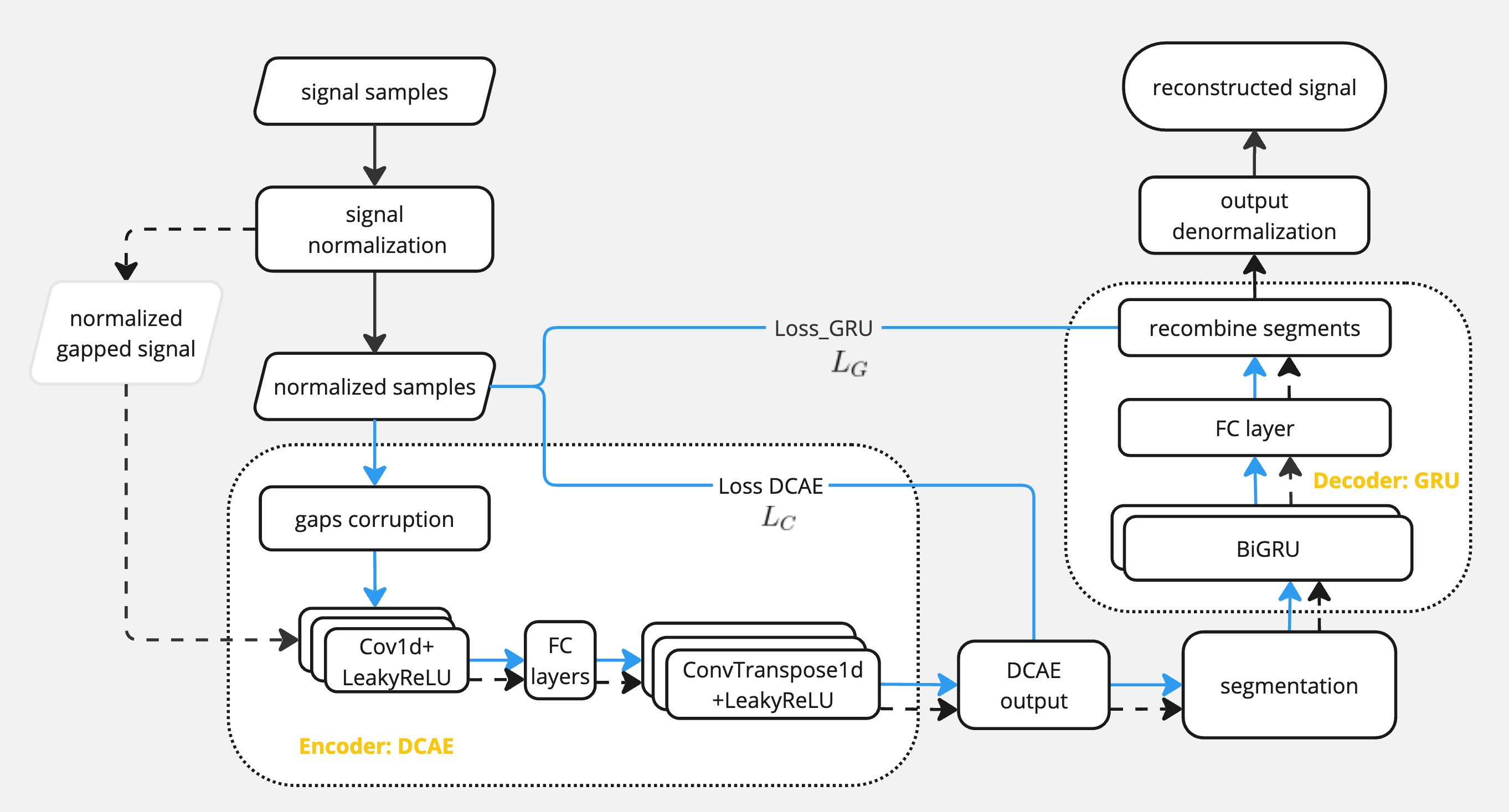}
    \caption{BiGRU-CAE model structure. The dashed box on the left illustrates the architecture of the DCAE, whereas the dashed box on the right depicts the structure of the BiGRU layers. The blue line represents the data flow for the training of the hybrid model with DCAE as the encoder and BiGRU layers in the decoder. The black dashed line represents the processing of the observed data stream with gaps in our proposed BiGRU-CAE model.}
    \label{fig:BiGRU-CAE}
\end{figure*}

\begin{figure}[h!]
    \centering
    \includegraphics[width=0.45\textwidth]{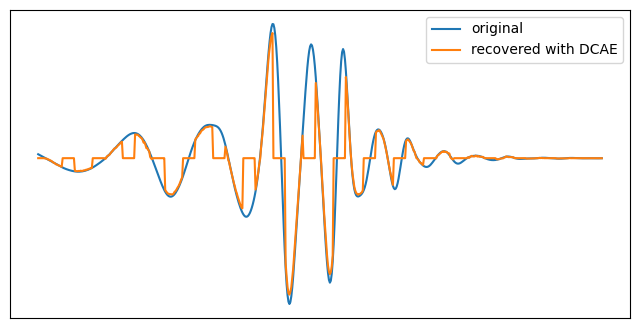}
    \caption{Example of DCAE outputs in the MBHB case. The blue line is a fraction of the merging phase of the test signal without noise. The orange line shows the according fraction of the output from the DCAE associated with the test signal with gaps. Discontinuities are discernible within the reconstructed signal, outlining the need for BiGRU layers.}
    \label{fig:DCAE}
\end{figure}

The structure of our model is depicted in Fig.~\ref{fig:BiGRU-CAE}. The data stream undergoes normalization before being used to train the DCAE. The left dashed box shows the training of the DCAE process. Complete normalized signals $d_{i}$ are corrupted with varying gap patterns for each iteration to ensure the robustness of the DCAE \cite{10.1145/1390156.1390294}. The detailed corruption process will be discussed in Section \ref{sec:data_prepare}. To address the issue of dying neurons associated with negative values, we employ the Leaky Rectified Linear Unit (LReLU) \cite{maas2013rectifier} as the activation function following each convolutional layer in the encoder and the transposed convolutional layer in the decoder. This choice facilitates an improved gradient flow and enhances the model's ability to learn robust feature representations. Due to the high dimension of the input, we did not apply the classic convolutional encoder-decoder architectures such as U-Net~\cite{ronneberger2015unet}, SegNet~\cite{7803544}, and DeepLabV3~\cite{chen2018encoderdecoder} for semantic image segmentation. In the output of the DCAE, minor isolated artifacts can be seen within the reconstructed signal $h_{C}(\hat{d_{i}})$ of corrupted signal $\hat{d_{i}}$; see the orange line in Fig.~\ref{fig:DCAE}.

To address these small corrupted fractions, we consider applying BiGRU layers to recover the continuity of the reconstructed signal. The training process can be seen with the right dashed box in Fig.~\ref{fig:BiGRU-CAE}. The input is the reconstructed signal after DCAE $h_{C}(\hat{d_{i}})$ with the newly corrupted signal. To make training computationally feasible, samples of these outputs are partitioned into several multiple subsequences of length $L$, where $L$ depends on the complexity of the input signal. The hidden states of size $K$ in the BiGRU are fed into the fully connected layer to predict the corrupted data. The observed signal with gaps $\Tilde{d_{i}}$ will follow the black line to produce the reconstructed signal after recombining the output of the BiGRU decoder $h_G(\hat{s}_{ij})$ and back normalization, where $\hat{s}_{ij}$ is the $j$-th segment of the DCAE output $h_C(\Tilde{d}_i)$.

\subsection{Data preparation}\label{sec:data_prepare}
In this subsection, we discuss the simulation of gaps in the data sequence. A binary mask function $w(t)$ is applied so that the simulated observed data with gaps is 
\begin{align}
\hat{d_{i}}=w(t) d_{i}
\end{align}
where 
\begin{equation}
w(t) = \begin{cases}
        1, & \text{if data at $t$ is available} \\
        0, & \text{if data at $t$ is unavailable}.
        \end{cases}
\end{equation}

The gaps are categorized into scheduled gaps and unscheduled gaps. Scheduled gaps result from periodic maintenance of the LISA spacecraft and its onboard instruments, such as antenna repointing, which typically causes predictable downtime of about 3.5 hours each week or longer interruptions of up to 7 hours every two weeks. In contrast, unscheduled gaps arise from unexpected hardware malfunctions or unforeseen physical occurrences, leading to unpredictable durations that can range from several hours to several days, significantly affecting the reliability of the collected data \cite{Dey_2021,Amaro_Seoane_2021}. Here, we will consider both scheduled and unscheduled gaps. The intervals between successive gaps are sampled from an exponential distribution, with its rate parameter defined according to the duty cycle 75\% requirements \cite{Dey_2021,10.1093/mnras/stab3314,Amaro_Seoane_2021}. The duration for the gaps is a uniform distribution between 4 and 8 hours per day and the total data sequence in our example is about 3 days. More flexible durations will be needed when considering longer signals.

\subsection{Training strategy}

Data normalization is a critical preprocessing step when training neural networks. The input data must be on a similar numeric scale to learn meaningful representations. Normalization transforms the input data stream to have zero mean and unit variance, ensuring all samples contribute equally to the learning process. This is typically achieved by subtracting the mean and dividing by the signal's standard deviation. Normalization helps prevent certain samples from dominating the reconstruction loss during training, which could lead to the autoencoder learning trivial or biased representations. 

Mean Squared Error (MSE) is commonly used as the loss function in a denoising autoencoder. The MSE measures the average squared difference between the autoencoder's predictions and the true, uncorrupted input values. This loss metric aims to minimize the discrepancy between the network's output and the original, unperturbed data, thereby enabling effective impainting. To mitigate overfitting for certain time periods or samples, L2 regularization is applied to discourage the autoencoder from learning large weight values. By adding a penalty proportional to the L2 norm of the weights, the \texttt{Adam} optimizer \cite{Kingma:2014} is encouraged to find a set of parameters with smaller magnitudes, resulting in a more generalized model.

Our BiGRU-CAE has two stacked autoencoders; a DCAE as the encoder, and an AE with BiGRU layers as the decoder, both of which have different loss functions. For DCAE, L2 regularization is added MSE, as discussed before to yield the following loss function: 
\begin{align}
L_C= \frac{1}{n}\sum_{i=1}^{n} (d_i - h_C(\hat{d}_i))^2+ \lambda \sum_{l=1}^{p} | \varphi_l |^2
\end{align}
where $h_C(\hat{d}_i)$ is an output of the DCAE with the corrupted samples; $\lambda$ is the L2 regularization hyperparameter, which is default to 0.001; $p$ is the total number of weight parameters in the autoencoder model; and $| \varphi_l |^2$ represents the L2 norm (Euclidean norm) of the $l$-th weight parameter $\varphi_l$. Data segmentation occurs before training of BiGRU layers. The loss function for the AE with a BiGRU is defined as:
\begin{align}
& L_G = \frac{1}{nJ}\sum_{i=1}^{n} \sum_{j=1}^{J}  (s_{ij} - h_G(\hat{s}_{ij}))^2\\
& h_C(\hat{d}_i) = \{\hat{s}_{i1},\hat{s}_{i2},...,\hat{s}_{iJ}\}
\end{align}
where $s_{ij}$ denote the $j$-th segment of length $L$ from the $i$-th complete sampled signal and, $J$ is the number of segments. After injecting simulated gaps that differ from those in the DCAE training, the DCAE output is partitioned into $J$-segments, denoted as $\hat{s}_{ij}$, $j=1,...,J$. The output of the hybrid model for each segment $h_G(\hat{s}_{ij})$ is stacked to form $h_{G}(h_{C}(\hat{d}_i))$.  Then the final reconstructed signal is obtained by denormalizing $h_{G}(h_{C}(\hat{d}_i))$. 
 Considering the overall pattern of the time sequence, the fractal Tanimoto similarity coefficient \cite{Diakogiannis_2021} can also be considered when dealing with complex signals \cite{Chatterjee_2021}.

\subsection{Bayesian theory}
Bayesian inference is the standard procedure used in GW astronomy to estimate parameters $\boldsymbol{\theta}$ given observations of a set of data streams $d_{o}$. At the heart of Bayesian theory lies Bayes' theorem: 
\begin{eqnarray}
    p(\boldsymbol{\theta}|d_{o}) &=& \frac{p(d_{o}|\boldsymbol{\theta})p(\boldsymbol{\theta})}{p(d_o)} \\
    &\propto& p(d_{o}|\boldsymbol{\theta})p(\boldsymbol{\theta}),
\end{eqnarray}
where $p(\boldsymbol{\theta}|d_{o})$ is the posterior density of unknown parameters $\boldsymbol{\theta}$ given the observation of a data stream $d_{o}$, $p(d_{o}|\boldsymbol{\theta})$ the likelihood function and $p(\boldsymbol{\theta})$ is the prior distribution, representing our knowledge about the parameters $\boldsymbol{\theta}$ before observing the data. The marginal likelihood $p(d_o) = \int_{\boldsymbol{\theta}\in\boldsymbol{\Theta}}p(d_{0}|\boldsymbol{\theta})p(\boldsymbol{\theta})\,\text{d}\boldsymbol{\theta}$ is a constant over the parameter space and is unnecessary in the parameter estimation of our experiment. 

Stochastic sampling algorithms, such as Markov chain Monte Carlo (MCMC), are used to obtain random samples $\boldsymbol{\theta}$ from the posterior density $p(\boldsymbol{\theta}|d_o)$ by constructing a Markov chain whose steady-state distribution is the posterior distribution of target parameter. In our work, we use an advanced MCMC sampler \texttt{Eryn}~\cite{Karnesis_2023}, which harnesses an ensamble affine-invariant sampler \cite{affine:2010} with parallel tempering to obtain samples from $p(\boldsymbol{\theta}|d_o)$ in the MBHB case.

The typical time-domain data stream observed by the LISA instrument will be a combination of TDI variables $X = \{A, E, T\}$, representing the response of the LISA instrument to the plus and cross polarisations of the incoming GW source in the transverse-traceless gauge~\cite{tinto:2021, tinto:2023}:
\begin{equation}\label{eq:data_stream_X}
d_{o}^{(X)}(t) = h_{\text{e}}^{(X)}(t;\boldsymbol{\theta}_{\text{0}}) + n^{(X)}(t), \quad X = \{A, E, T\}.
\end{equation}
Here $d_{o}$ is the observed data stream, $\boldsymbol{\theta}_{0}$ are the true parameters of the true gravitational wave $h^{(X)}_{\text{e}}$, and $n^{(X)}(t)$ are noise fluctuations arising from perturbations to the LISA instrument from unresolvable GW sources and non-GW instrumental perturbations. In this paper, to explain the whole imputation procedure, we take the data stream on channel A. Under the assumption that the noise is stationary and follows a Gaussian distribution in the parameter estimation process, the log-likelihood with inner product \footnote{
\begin{equation}\label{eq:inner_prod}
    (a|b)^{(X)} = 4\text{Re}\int_{0}^{\infty}\text{d}f\frac{\hat{a}^{(X)}(f)(\hat{b}^{(X)}(f'))^{\star}}{S^{(X)}_{n}(f')}.
\end{equation}
where $S_n^{(X)}$ is the power spectral density (PSD) of the noise process within a channel $X$. Hatted quantities refer to the Fourier transform with convention,
\begin{equation}
    \hat{h}(f) = \int_{0}^{\infty}\text{d}t \, h(t)\exp(-2\pi \text{i} f t).
\end{equation}
} ~\cite{finn1992detection, Flanagan:1997kp} is
\begin{equation}\label{eq:whittle_likelihood_signal}
p(d|\boldsymbol{\theta}) = -\frac{1}{2}\sum_{A}(d - h_m|d - h_m)^{(A)},
\end{equation}
where $h_m$ are model templates, generating the likelihood when inferring parameters $\boldsymbol{\theta}$ with MCMC. Window functions are employed to reduce the effects of spectral leakage in Fourier transform analysis, which are then utilized in the following section on the toy model case as a smoothing taper to formulate the likelihood of the signal with gaps.

\section{Application}
In this section, we demonstrate how to impute the gaps with our proposed model in a simple toy example in Section \ref{sec:toy_model} and a MBHB signal in Section \ref{sec:massive_black_holes_application}. Noise is not included here to examine the performance of the proposed model for imputation only. This analysis assumes a denoising procedure has been applied earlier in the pipeline, and this will be the focus of forthcoming research. The simulated data are partitioned into training and validation sets, with 80\% of the samples allocated for training and 20\% reserved for validation.

\subsection{Toy model case} \label{sec:toy_model}

Consider a data stream of the form
\begin{equation}\label{eq:data_stream_toy}
d(t;a,f,\dot{f}) = a\sin \left(2\pi t\left[f + \frac{1}{2}\dot{f}t\right]\right).
\end{equation}

Assume an observed test signal $d_o$ with parameters $\boldsymbol{\theta}_0 = \{a_{0} = 5\cdot 10^{-21}, f_{0} = 10^{-3}\,\text{Hz}, \dot{f}_{0} = 10^{-8}\,\text{Hz}/\text{s}\}$. The training set contains 800 generated signals using prior samples.
\begin{align*}
    a &\sim \text{U}[a_{0}-10^{-21}, a_{0}+10^{-21}]\\
    f &\sim \text{U}[f_{0}-10^{-6}, f_{0}+10^{-6}]\,\text{Hz}\\
    \dot{f} &\sim \text{U}[\dot{f}_{0}-10^{-12}, \dot{f}_{0}+10^{-12}]\, \text{Hz}/\text{s}. 
\end{align*}
Here, we simulated the signal with an observation period of 3 days sampled with cadence $\Delta t = 5$ seconds, yielding a signal length of 51480. SNRs determined by equation (8) of ~\cite{PhysRevD.109.083002} are between 80 and 200.

The structure of DCAE in the proposed model is in Table \ref{tab:autoencoder-structure}. Due to memory limitations, 160 outputs of DCAE will be sampled and cut into pieces with a length of 48 due to a simple signal structure. In the training of the BiGRU component, complete signal samples will first pass through the DCAE and then pass through a two-layer bi-directional gated recurring unit with a hidden state size $K=12$ followed by a fully-connected layer and a \texttt{Tanh} activation function. The training time is about $0.5$ hours for DCAE with 100 epochs and $2.5$ hours for AE with BiGRU layers with 50 epochs. The validation loss can be seen in Fig.\ref{fig:combined_loss_toy} in Appendix \ref{Appendix:A}.

\begin{table*}
\centering
\caption{\small DCAE Model Structure.}
\label{tab:autoencoder-structure}
\begin{tabular}{|l|c|c|c|c|c|}
\toprule
\hline
Layer Type & Input Channels & Output Channels & Kernel Size & Stride \\
\hline
\midrule
Conv1d     & 1                   & 16                    & 7           & 1        \\
Conv1d     & 16                  & 32                    & 5           & 4        \\
Conv1d     & 32                  & 64                    & 3           & 8        \\
Flatten    & 64          & 103680                & -           & -                 \\
Linear     & 103680              & 4096                  & -           & -        \\
Linear     & 4096                & 1024                  & -           & -       \\
Linear     & 1024                & 4096                  & -           & -       \\
Linear     & 4096                & 103680                & -           & -      \\
Unflatten  & 103680              & 64             & -           & -       \\
ConvTranspose1d & 64             & 32                    & 3           & 8      \\
ConvTranspose1d & 32             & 16                    & 5           & 4       \\
ConvTranspose1d & 16             & 1                     & 7           & 1       \\
\bottomrule
\hline
\end{tabular}
\end{table*}

\begin{figure*}
    \centering
    \includegraphics[width=0.9\textwidth]{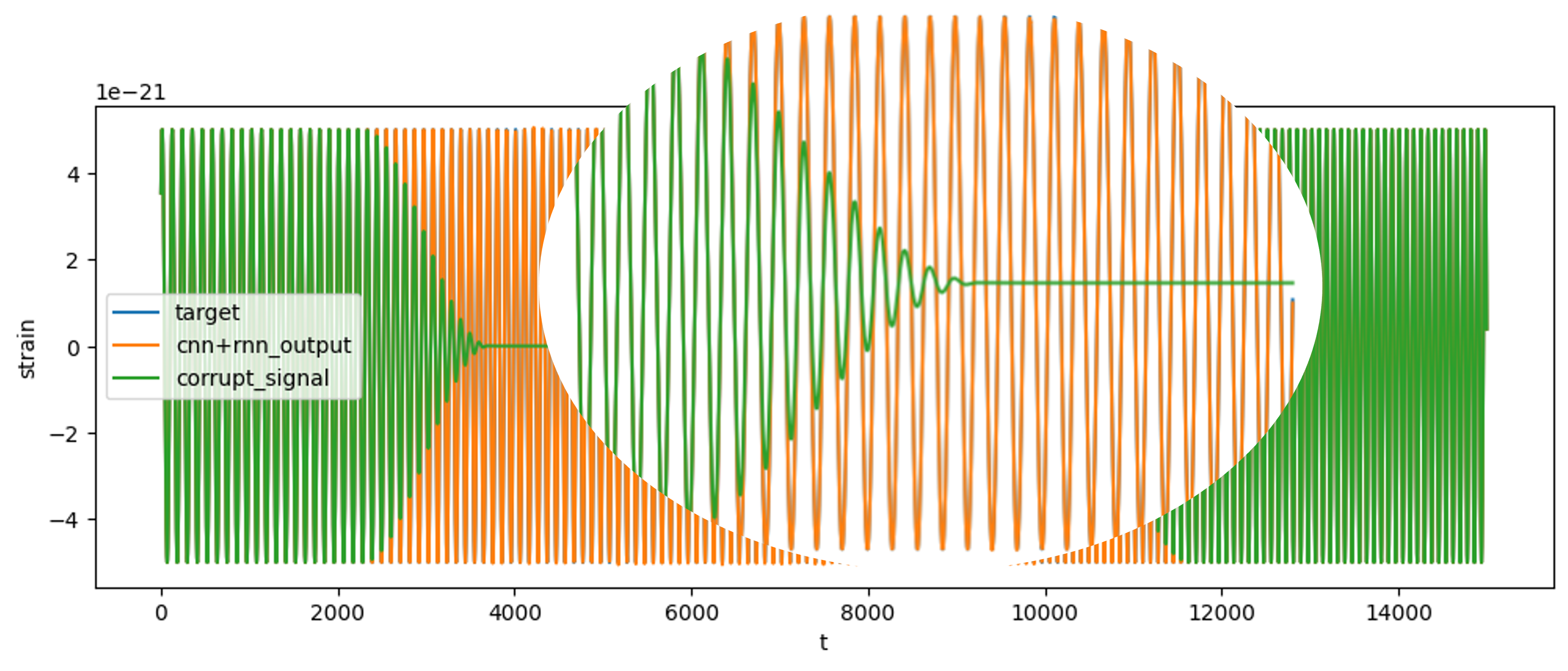}
    \caption{Segment for example of imputing the toy model signal. The green line represents the corrupted signal with the window function. The blue line (target) is the true shape of the test signal. The orange line represents the final output of the proposed model which combines the DCAE and BiGRU. The output orange line completely masks the target blue line, rendering the target blue line nearly invisible.}
    \label{fig:result_model}
\end{figure*}

Parameter estimation via MCMC will be considered on the back-transformed output of the proposed model. We tested the signal with different gaps defined in Section \ref{sec:data_prepare} and compared their posterior distributions $p(\boldsymbol{\theta}|h_{G}(h_{C}(\hat{d}_i))$ with the original one $p(\boldsymbol{\theta}|d_i)$. The process of the Bayesian parameter estimation is similar to the toy model case in \cite{PhysRevD.109.083002}.  A test case characterized by an on-duty percentage of 87.5\% was utilized as a representative example. The proposed model exhibits a superior level of accuracy in estimating the missing data within the gaps, as demonstrated by the output orange line entirely overlapping the target blue line in Fig.~\ref{fig:result_model}. 

\begin{figure*}
    \centering
    \includegraphics[width=0.8\textwidth]{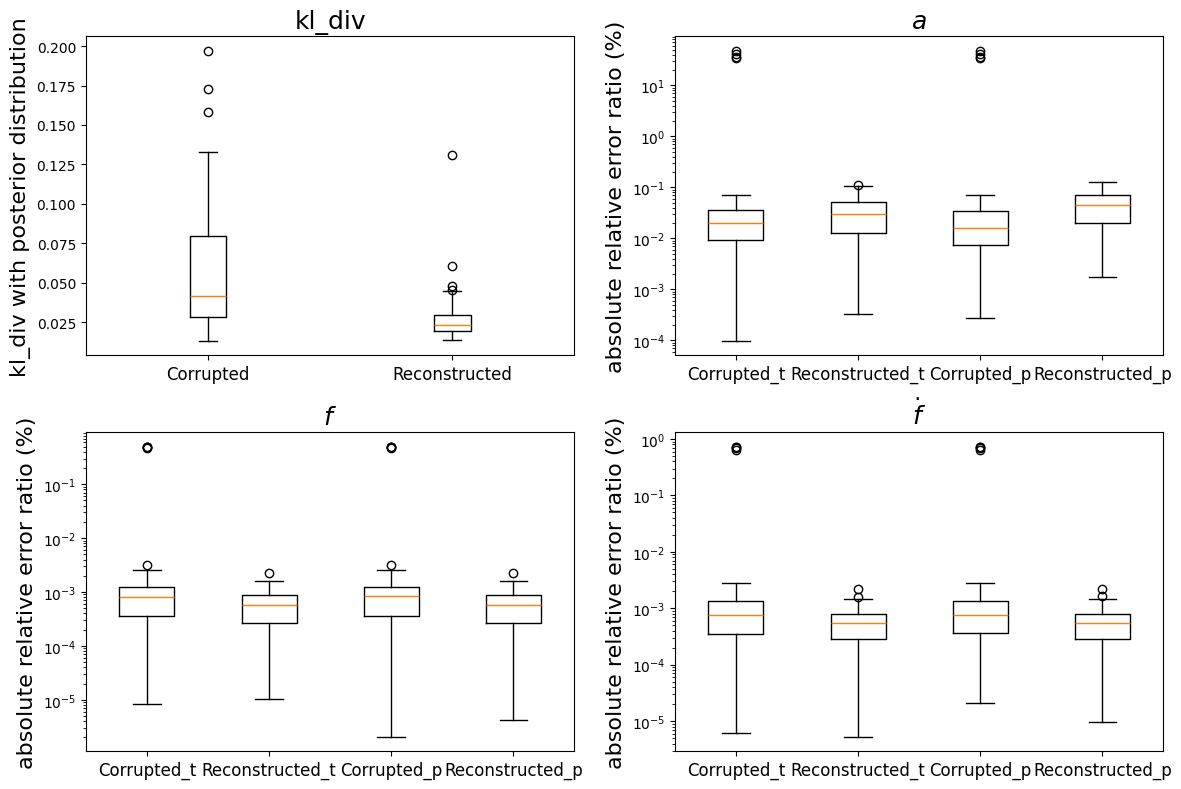}
    \caption{Result of the parameter estimation for the toy model signal. ``Corrupted" and ``Reconstructed" represent the signal with gaps and the reconstructed signal; Subscript ``t" represents the results compared with the true value; Subscript ``p" represents the result compared with the posterior mean of the original distribution of the complete signal.}
    \label{fig:result_model_p}
\end{figure*}

We consider 85 different scenarios where the test signal is corrupted with gaps. The results of parameter estimation are demonstrated in Fig.~\ref{fig:result_model_p}, which provides a visual comparison of the Kullback-Leibler (KL) divergence \footnote{The Kullback-Leibler divergence of $p_1(x)$ from $p_2(x)$ is $D_{KL}(p_1(x)\|p_2(x))=\mathbb{E}_{p_1(x)}[\log p_1(x)-\log p_2(x)]\approx \frac{1}{m}\sum^m_{i=1} \log p_1(x_i)-\log p_2(x_i)$ where $x_i\sim p_1$.} and absolute relative error for reconstructed signals and corrupted signals with gaps, to assess the performance of parameter estimation based on reconstructed signals relative to the true values and the posterior distribution of the signal without gaps. Since the exact waveform is known, the window function is applied in the Bayesian inference process for the signal with gaps, called ``corrupted" signals. Let $p(\boldsymbol{\theta}|h_{G}(h_{C}(\hat{d}_o)))$ be the posterior of the imputed signal and $p(\boldsymbol{\theta}|d_o)$ be the posterior of original signal. The Gaussian kernel density estimation with the bandwidth selected by Scott’s rule \cite{scott2015multivariate} is denoted by $\hat{p}$. The summary of KL divergence $D_{KL}(\hat{p}(\boldsymbol{\theta}|d_o)||\hat{p}(\boldsymbol{\theta}|h_{G}(h_{C}(\hat{d}_o))))$ and $D_{KL}(\hat{p}(\boldsymbol{\theta}|d_o)||\hat{p}(\boldsymbol{\theta}|\hat{d}_o))$ from 85 replicates are shown the top-left plot in Figure \ref{fig:result_model_p}. 

In general, a higher median and wider spread of KL divergences for signals with gaps are observed, even though 4 cases of infinite values were ignored in the plot. The boxplots with log scale comparing the absolute relative error ratio show our proposed model's ability to mitigate large biases when doing Bayesian inference with corrupted signals. One example of this scenario is displayed in Fig.\ref{fig:result_model_e} in Appendix \ref{Appendix:B}. This result shows that, on average, the reconstruction signal was more useful in parameter estimation than using the corrupted signal. 

\subsection{Massive black hole binary case} \label{sec:massive_black_holes_application}
In this section, we tested our proposed model on a realistic massive black hole binary signal with unscheduled gaps defined in section \ref{sec:data_prepare}. Using \texttt{LISA Analysis Tools} \cite{michael_katz_2024_10930980}, we generate and then analyze complete inspiral-merger-ringdown frequency domain spin-aligned MBHBs in the solar system barycenter frame with the LISA response applied. The training set contains 4,000 samples generated by \texttt{IMRPhenomHM} waveforms and \texttt{BBHx} \cite{Katz_2020,Katz_2022,Khan_2016,London_2018,Husa_2016} with a uniform distribution on the primary mass $m_{1} \sim \text{U}[1.5 \times 10^6,2.5\times 10^6]$. The data stream of channel A is transformed into the time domain to implement the proposed model. The structure of the DCAE component is similar to that of the toy model in Table \ref{tab:autoencoder-structure}, while we cut the sequence into pieces with a length of 1024 before training the BiGRU component, which is longer than in the case of the toy model with a simpler signal structure. The training time is about $4.3$ hours when training DCAE with 100 epochs and $3$ hours when training BiGRU with 50 epochs since the number of sampled outputs of DCAE is still 160. The validation loss can be seen in Fig.\ref{fig:combined_loss_mbhb} in Appendix \ref{Appendix:A}.

\begin{figure*}
    \centering
    \includegraphics[width=0.9\textwidth]{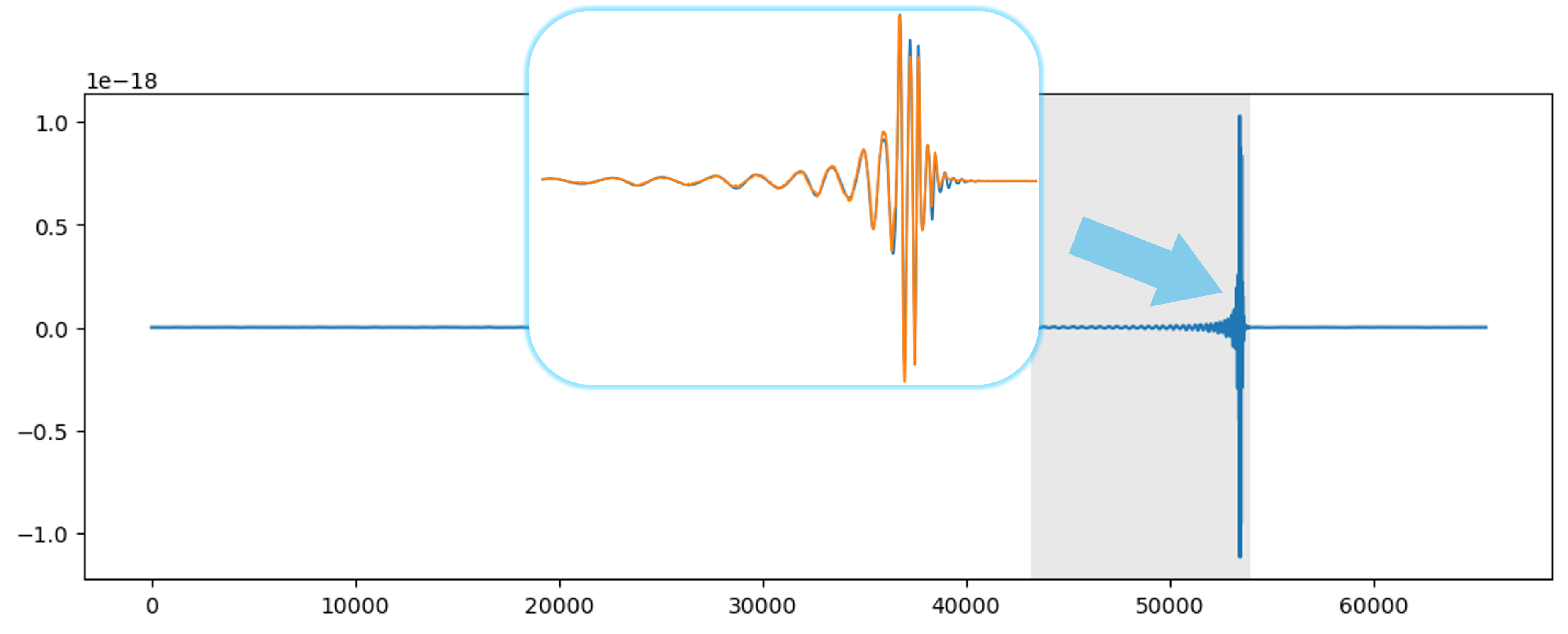}
    \caption{Example of the imputation results for the corrupted signal for a MBHB signal. The grey area depicts a section where a long-duration gap happened. The orange line in the zoomed picture represents the imputation result through the proposed model.}
    \label{fig:result_s_gap}
\end{figure*}

To test the performance of our model, we run MCMC to do parameter estimation focusing on a subset of parameters $\boldsymbol{\theta} = \{m_T, q,\phi_{\text{ref}}\}$. The \texttt{Heterodyned} likelihood is used to speed up the Bayesian inference process \cite{zackay2018relative}. The parameters for the test signal are defined as follows: the total mass $m_T = m_{1} + m_{2} = 2.7\cdot 10^{6}M_{\odot}$; mass ratio $q = m_{1}/m_{2} = 0.35$; the two effective spin parameters of the two-component masses $\chi_{1} = 0.5$ and $\chi_{2} = 0.7$; the reference time $t_{c} = 10^{6}\,$seconds; luminosity distance $15\,$ Gpc; the reference phase $\phi_{\text{ref}}=0.6$; sky position $(\beta = 0.7, \lambda = 3.4)$ in ecliptic coordinates; and polarisation angle $\psi = \pi/4 $. The observation time will be $\sim 3$ days, sampled with cadence $\Delta t = 4$ seconds. Therefore, the length of the data sets is $N= 65536$ with an SNR of 3233.49. The basic LISA sensitivity is applied. Fig.~\ref{fig:result_s_gap} illustrates the imputation results for the corrupted signal. Despite the substantial loss of information, especially during the merger period, the proposed model effectively captures the principal characteristics of the sequence, as evidenced by the high degree of similarity between the orange line representing the recovery signal and the blue line denoting the target original signal.

\begin{figure*}
    \centering
    \includegraphics[width=0.8\textwidth]{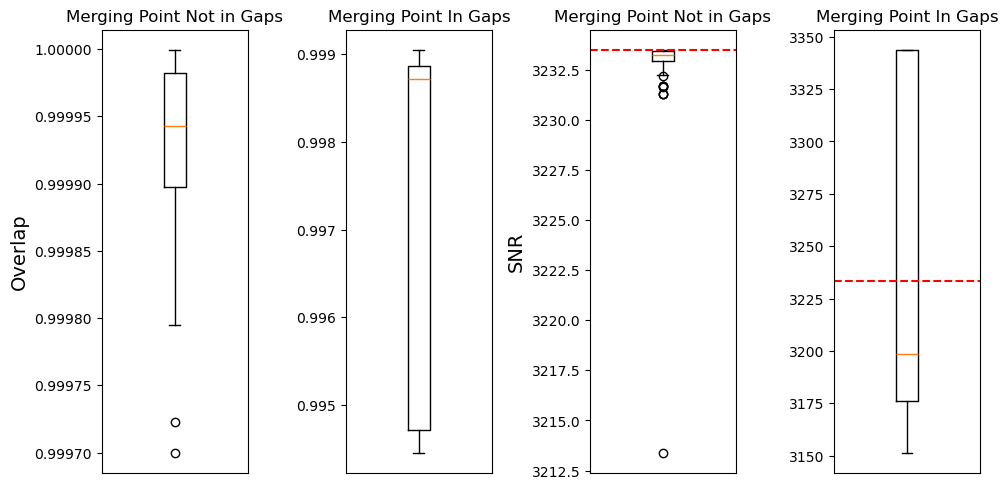}
    \caption{Boxplot of overlap (first and second plots) and SNR (third and fourth plots) for the massive black hole test signal. Different scales are used in boxplots for merging point not in gaps and merging point in gaps. The red dashed lines in the third and fourth plots represent the true SNR.}
    \label{fig:result_s_o}
\end{figure*}

We consider 86 different cases where the test signal is corrupted with gaps. We then calculated the SNR and overlap of the recovered signals \footnote{The overlap is defined as the inner product of the complete signal and the reconstructed signal: 
\begin{align} O(d_o,h_G(h_c(\hat{d}_o))) = <d_o|h_G(h_c(\hat{d}_o))> \end{align}}. Results can be seen in Fig.~\ref{fig:result_s_o}. 

\begin{figure*}
    \centering
    \includegraphics[width=0.8\textwidth]{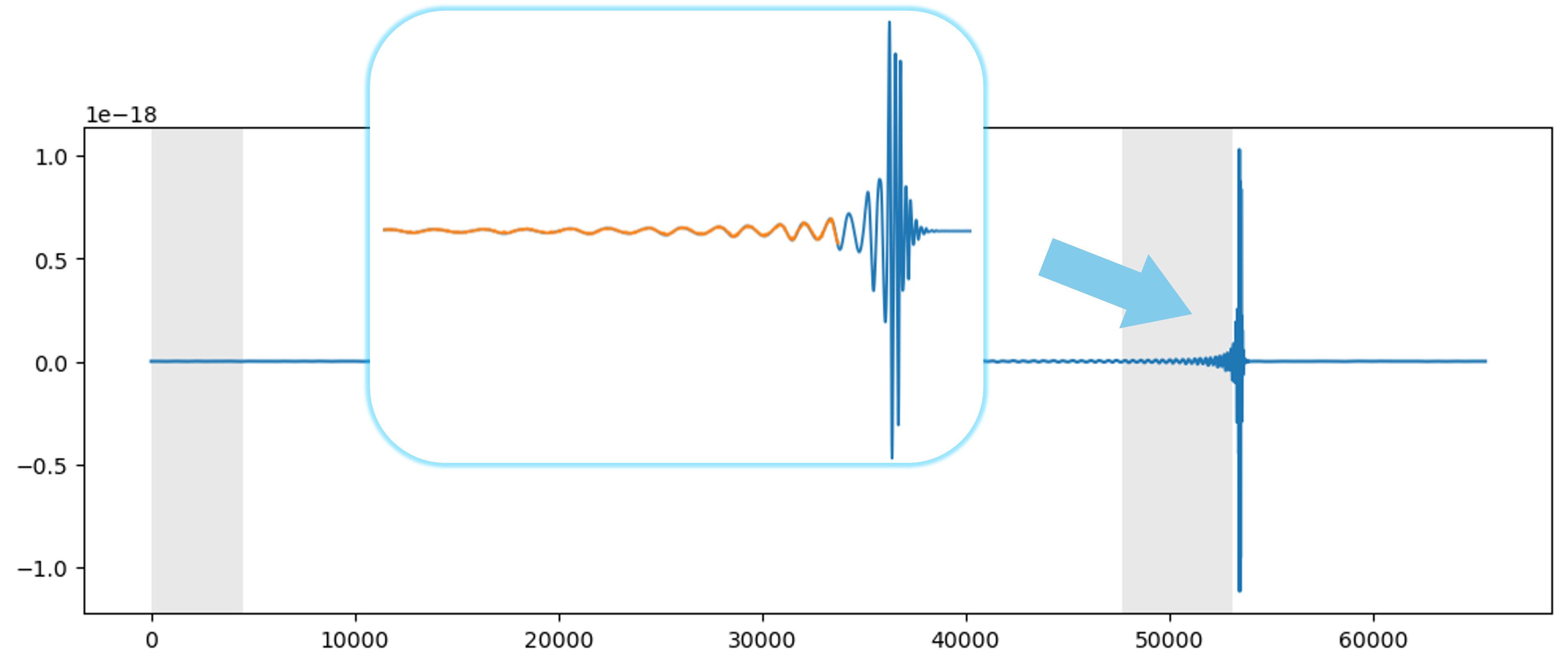}
    \caption{Example for gaps near merging point for massive black signal. Same setting as Fig.~\ref{fig:result_s_gap}}
    \label{fig:example_merging}
\end{figure*}

All of the recovered signals achieve an overlap larger than 99\%, and their SNRs are close to the original SNR. However, the results of the recovered signals with merging occurring in gaps differ significantly from those with gaps that do not contain the merging point. Two outliers in the overlap metric are observed for signals where the merging does not occur in gaps. One outlier corresponds to a substantial unduty cycle of 35.7\%, while the other arises due to a gap in the signal occurring near the merging point, see Fig.~\ref{fig:example_merging}, which also contributes to the outlier in SNR. This observation aligns with the findings reported in \cite{Dey_2021}, which demonstrate a greater impact as the gaps approach the merger point. Nevertheless, this effect is attenuated by our model during parameter estimation, as evidenced by the insignificant bias depicted in Fig.~\ref{fig:result_s_n} in Appendix \ref{Appendix:B}.

\begin{figure*}
    \centering
    \includegraphics[width=0.8\textwidth]{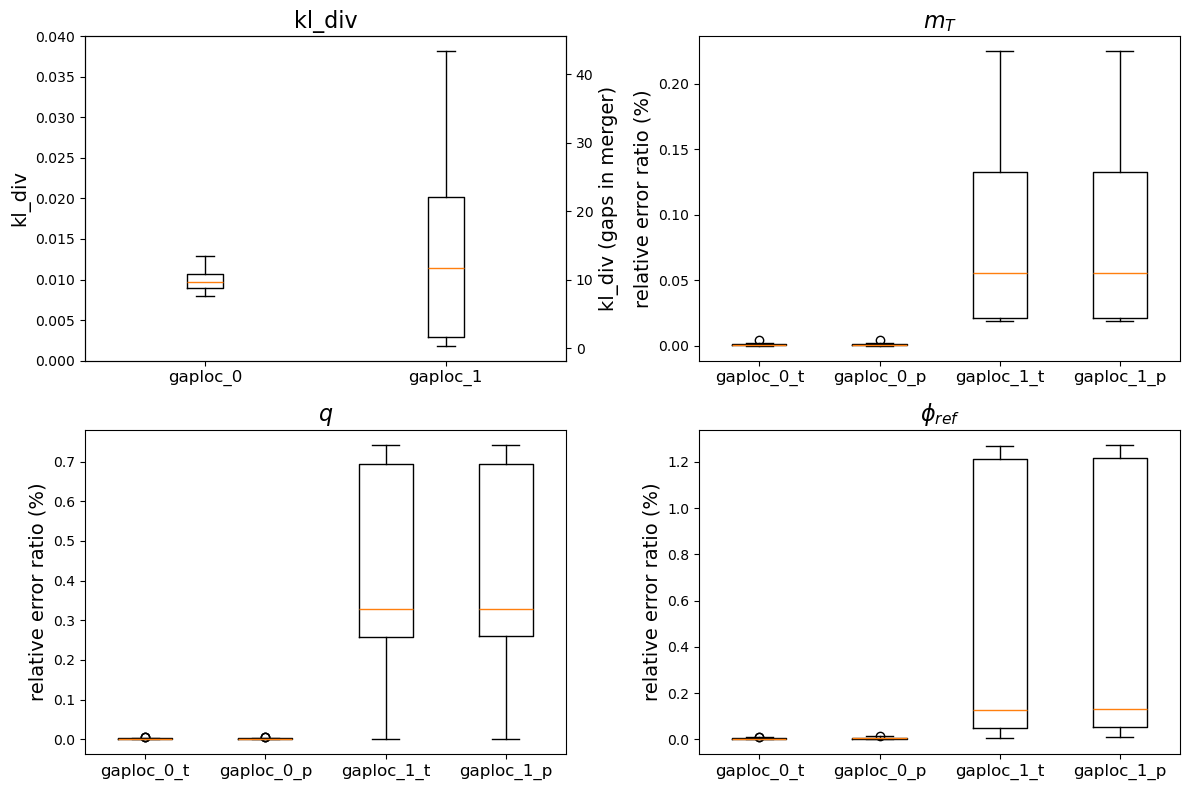}
    \caption{Boxplots of KL divergence and relative error for parameter estimation for massive black hole signal. Subscripts 0 and 1 following ``gaploc" indicate gap occurrence not in and in the merging phase respectively.
     Subscripts ``t" and ``p" indicate the posterior mean comparison to the true value and posterior mean using the original complete signal, respectively. 
    }
    \label{fig:result_k_e}
\end{figure*}

Likewise, we conducted an analysis of the performance of the reconstructed signal in parameter estimation. Given the unavailability of the precise waveform in the time domain and the inherent computational challenges when performing Bayesian inference on the ``corrupted" signals, our investigation is confined to reconstructed signals. The parameter estimation for signals with no gaps at the merging point performs better than those with gaps at the merging point. In Fig.~\ref{fig:result_k_e}, the absolute relative error ratios are less than 1.3\% for all three parameters, and the bigger relative error ratio for unscheduled gaps at the merger shows some limitations of our signal reconstruction method. 
The boxplot for KL divergence shows even a bigger disparity in uncertainty estimation for parameters, notwithstanding the exclusion of seven infinite cases and one case exceeding 150 for gaps occurring during the merger. When unscheduled gaps happened at merger time, the gaps were poorly reconstructed, and KL divergences were 1000 times bigger on average. This is observed in the posterior comparison for the two cases in Fig.~\ref{fig:result0} and Fig.~\ref{fig:result1} in Appendix \ref{Appendix:B}. 
This underscores the necessity to investigate scenarios in which scheduled gaps occur during the merging phase.

\begin{figure*}
    \centering
    \includegraphics[width=0.8\textwidth]{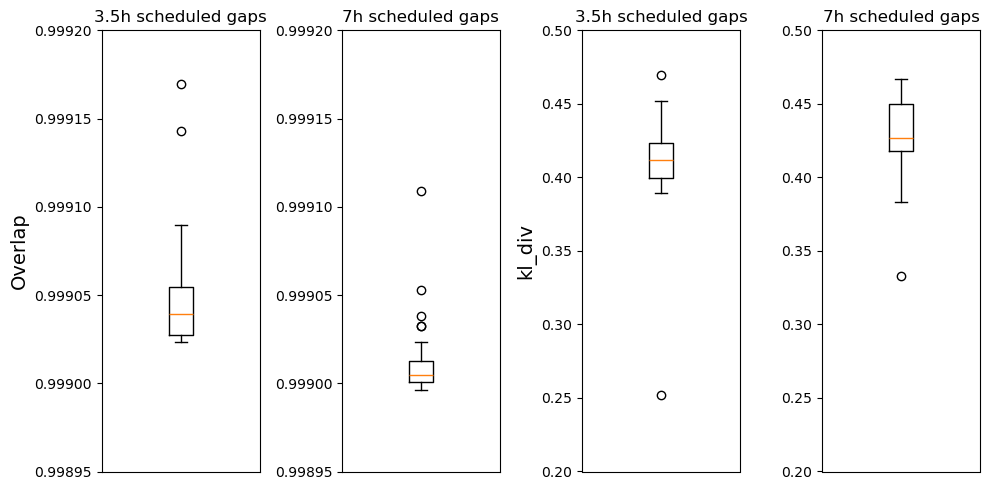}
    \caption{Boxplots of overlap (first and second plots) and KL divergence (third and fourth plots) for posterior distribution in parameter estimation for massive black hole signal with $3.5$-hour and $7$-hour scheduled gaps in the merger. }
    \label{fig:result_she}
\end{figure*}

To investigate this issue, a total of 60 distinct instances of scheduled gaps were analyzed, comprising 30 instances of 3.5-hour gaps and 30 instances of 7-hour gaps, occurring during the merging time with various injection times. It is not surprising to observe that the parameter estimates of these reconstructed signals exhibit some deviations from the true posterior distribution, as illustrated in Fig.~\ref{fig:result35} and Fig.~\ref{fig:result7} in Appendix \ref{Appendix:B}. In comparison to unscheduled gaps occurring during mergers, reconstructed signals with scheduled gaps exhibit notably fewer distortions as measured by the KL divergence, less than 0.5. This observation is also consistent with the findings reported in \cite{Dey_2021}. Furthermore, the analysis indicates that the reconstructed signals associated with 7-hour scheduled gaps exhibit a slightly greater impact for parameter estimation than 3.5-hour scheduled gaps, as it is shown in Fig.~\ref{fig:result_she}. This highlights the need for protected periods within the LISA data stream, particularly when adopting a biweekly maintenance schedule, to prevent scheduled gaps from coinciding with the merger phase of a signal. Therefore, to mitigate possible biases in parameter estimates, we recommend not scheduling antenna repointing during merger time.

\section{Discussion}\label{sec:discussion}
In this paper, we have proposed an innovative BiGRU-CAE hybrid model that leverages the strengths of both convolutional autoencoders and gated recurrent units to address the challenges posed by data gaps in LISA gravitational wave observations. The DCAE component is well-suited for extracting relevant features from the high-dimensional input signals, while the BiGRU component can effectively capture the temporal dynamics of the gravitational wave signals. With the limitation of computation,  our research focuses on gaps in data streams without noise on one channel. The toy model study demonstrates improvements in parameter estimation when applying our model. It is noteworthy to examine the influence of the timing of data gaps in the context of the realistic massive black hole case study. This underscores the imperative of optimizing the maintenance schedule of the interferometer to mitigate potential biases in gravitational wave (GW) analysis that could result from data gaps.

One key advantage of this hybrid approach is its ability to handle long-duration and interrupted LISA data streams. The convolutional layers within the DCAE are adept at efficiently processing the input signals, while the GRU component is proficient in modeling temporal correlations 
to ameliorate discontinuities present in the output of the DCAE. This is a significant improvement over previous methods, which struggled with the computational costs and modelling errors associated with handling long gaps. The end-to-end training of the BiGRU-CAE model simplifies subsequent analysis steps compared to traditional Bayesian augmentation methods, which treat the missing data as auxiliary variables \cite{Baghi_2019}. Instead of chopping the data stream into pieces before the training of the autoencoder \cite{Xu:2024jbo}, our model trains the autoencoder with the whole sequence, which gives more robustness when imputing the gaps. 

Separating a complex deep-learning model into two stacked components allows for greater modularity and flexibility. Each component can be developed, trained, and optimized independently, making the overall model more adaptable and easier to iterate. It also reduces the overall computational requirements. Furthermore, each component can be designed and trained to specialize in a specific task or learn a particular set of features, which improves performance compared to a single, monolithic model that has to learn all the necessary capabilities.

This study constitutes the preliminary investigation of the gaps at merger in signals from MBHB. Although our model exhibits competence in recovering signals with gaps not located at the merger point, minor discrepancies from the true values in the reconstructed signal with gaps at merger suggest that further research is needed. While we opted for a simplified scenario for the pure signal to illustrate the applicability of our method, subsequent research should incorporate noise into the data stream, thereby enabling the architecture to perform denoising and imputation tasks simultaneously. This approach will align more closely with the realistic conditions encountered in current LISA data processing. In addition, certain computational constraints still exist. We plan to retrain our proposed model utilizing GPU with larger training dataset to process long data streams on the A and E channels, employing a 2D convolutional autoencoder. As our model is predominantly signal-focused, our plan is to develop a more efficient denoising and inpainting pipeline capable of processing signals from diverse sources in the future.

The \texttt{Python} code for the toy model case in Section \ref{sec:toy_model} is provided at \url{https://github.com/bpandamao/BiGRU_CAE}. 

\section*{Acknowledgements}
We thank Ollie Burke for his helpful discussions. All computations are performed on a virtual machine with 32GB RAM, 16 VCPUs, and an Ubuntu Linux operating system. The autoencoder and bi-directional gated recurrent unit were implemented using Python package \texttt{PyTorch}. We thank the Center for eResearch (CeR) at the University of Auckland for providing access to and assistance with the Nectar Research Cloud. Ruiting Mao would like to thank the University of Auckland for a UoA Doctoral scholarship. MCE and JEL acknowledge support by the Marsden grant MFP-UOA2131 from New Zealand Government funding, administered by the Royal Society Te Aparangi. 

\clearpage  

\onecolumngrid  

\appendix

\section{Loss during Training in toy model and MBHB case}\label{Appendix:A}

\begin{figure}[h]
    \centering
    \subfloat[Validation Loss each epoch for DCAE]{%
        \includegraphics[width=0.55\textwidth]{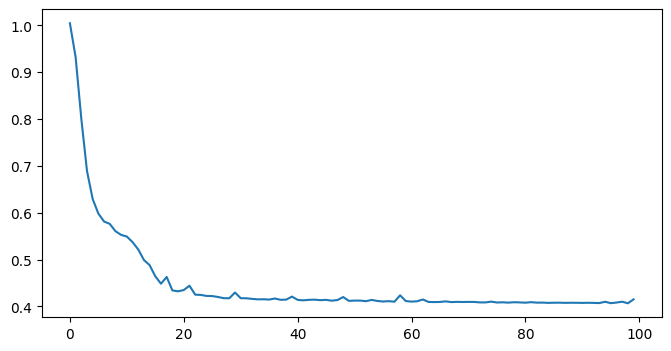} 
        \label{fig:loss_dcae_toy}
    }
    \hfill
    \subfloat[Validation Loss each epoch for BiGRU]{%
        \includegraphics[width=0.4\textwidth]{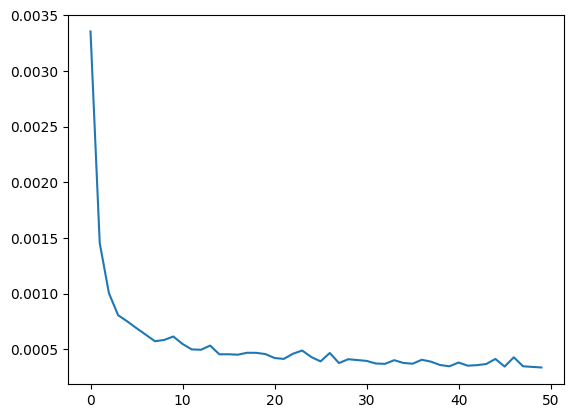} 
        \label{fig:loss_gru_toy}
    }
    \caption{Validation Loss when training with toy model}
    \label{fig:combined_loss_toy}
\end{figure}

\begin{figure}[h]
    \centering
    \subfloat[Validation Loss each epoch for DCAE]{%
        \includegraphics[width=0.55\textwidth]{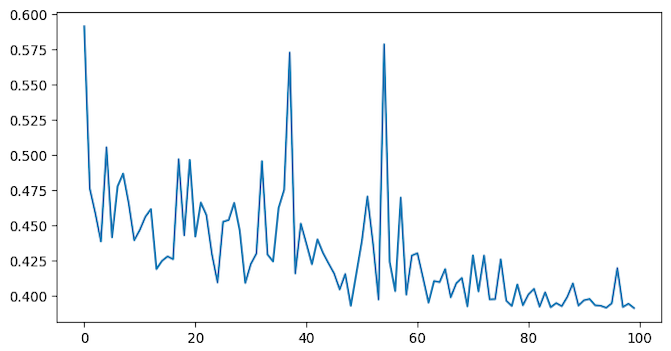} 
        \label{fig:loss_dcae_mbhb}
    }
    \hfill
    \subfloat[Validation Loss each epoch for BiGRU]{%
        \includegraphics[width=0.4\textwidth]{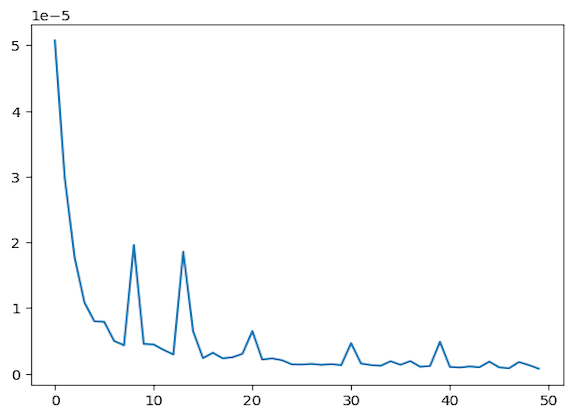} 
        \label{fig:loss_gru_mbhb}
    }
    \caption{Validation Loss when training with MBHB}
    \label{fig:combined_loss_mbhb}
\end{figure}

\pagebreak

\section{Examples of the parameter estimation in application}\label{Appendix:B}

\begin{figure}[h]
    \centering
    \includegraphics[width=0.8\textwidth]{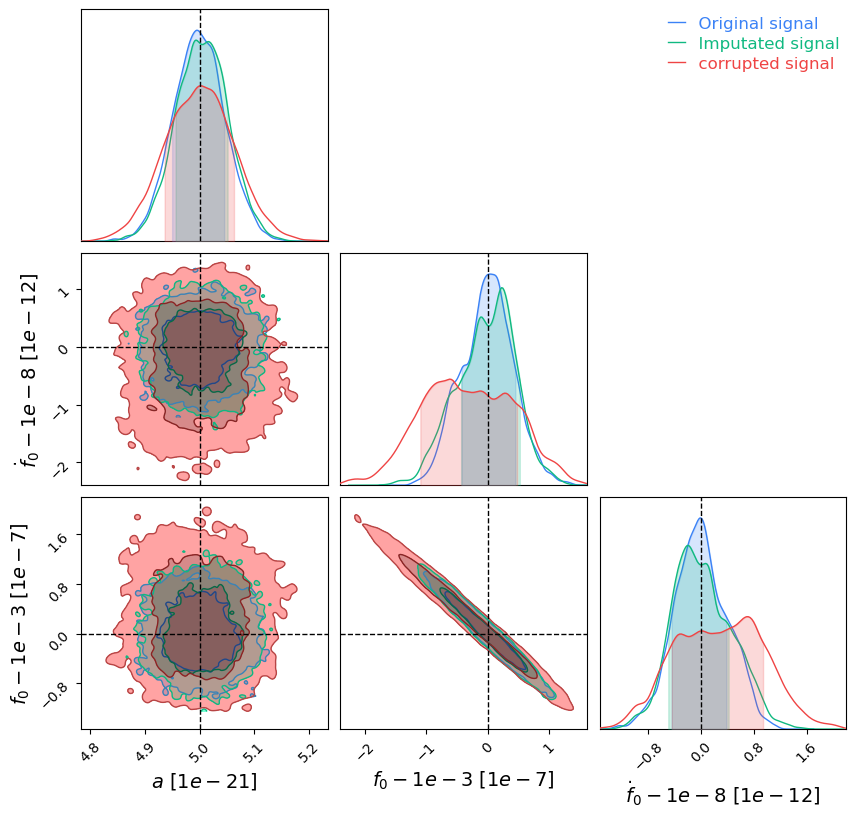}
    \caption{Example in a toy model of comparison of the posterior distribution among original, corrupted, and reconstructed signal. The black dashed line represents the true value. The shadowed areas represent the posterior incredible interval for one error in marginal distributions.}
    \label{fig:result_model_e}
\end{figure}

\begin{figure}
    \centering
    \includegraphics[width=0.8\textwidth]{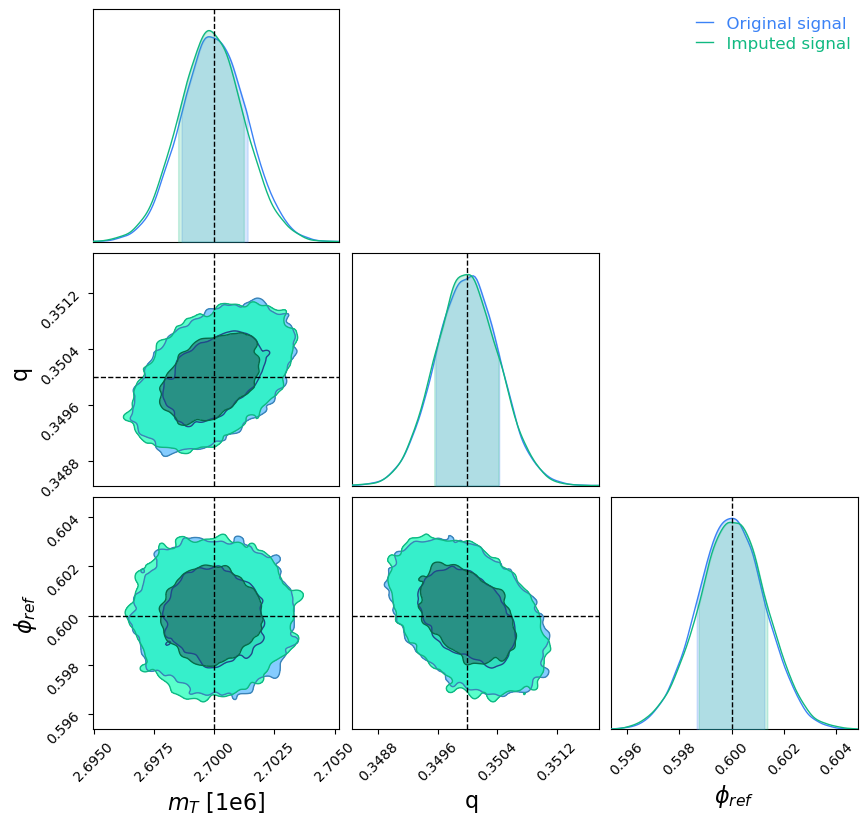}
    \caption{Parameter estimation example in MBH case for gaps near merging point. The black dashed line represents the true value. The shadowed areas represent the posterior incredible interval for one error in marginal distributions.}
    \label{fig:result_s_n}
\end{figure}

\begin{figure}
    \centering
    \includegraphics[width=0.8\textwidth]{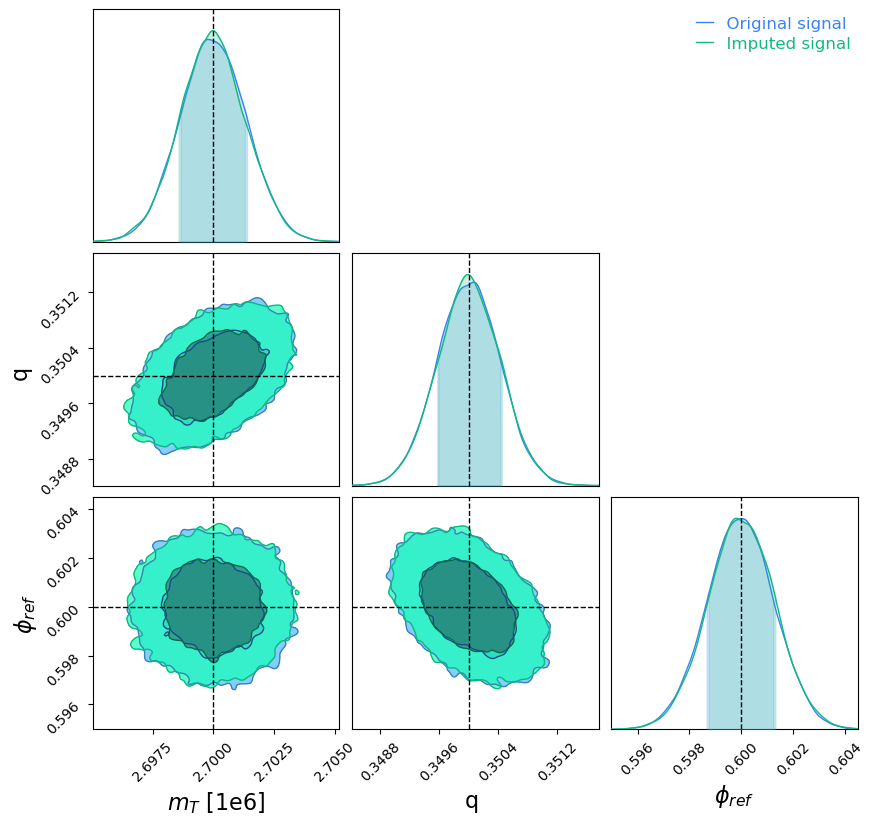}
    \caption{Parameter estimation example in MBH case for gaps not in merging time. The black dashed line represents the true value. The shadowed areas represent the posterior incredible interval for one error in marginal distributions.}
    \label{fig:result0}
\end{figure}

\begin{figure}
    \centering
    \includegraphics[width=0.8\textwidth]{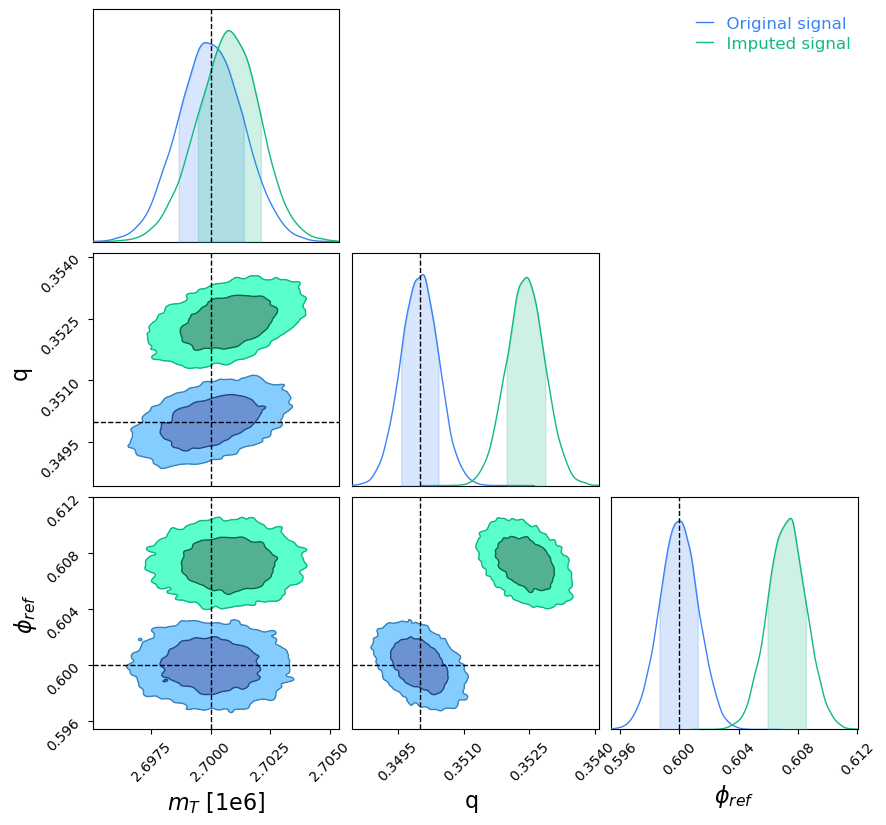}
    \caption{Parameter estimation example in MBH case for gaps in merging time. The black dashed line represents the true value. The shadowed areas represent the posterior incredible interval for one error in marginal distributions.}
    \label{fig:result1}
\end{figure}

\begin{figure}
    \centering
    \includegraphics[width=0.8\textwidth]{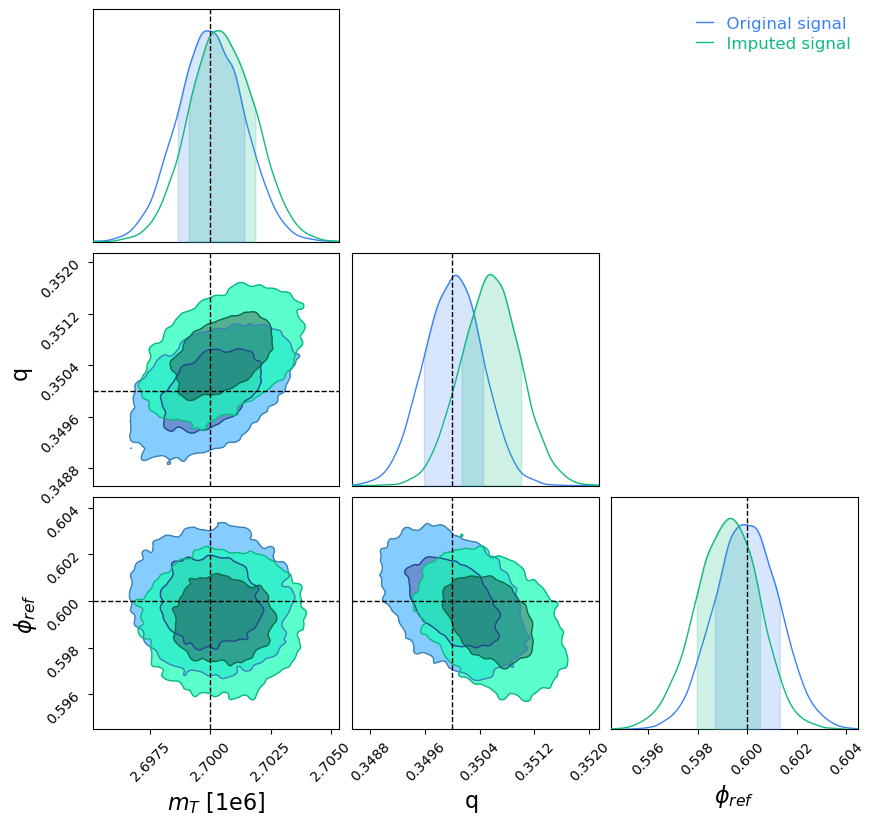}
    \caption{Parameter estimation example in MBH case for $3.5$-hour scheduled gaps in merging time. The black dashed line represents the true value. The shadowed areas represent the posterior incredible interval for one error in marginal distributions.}
    \label{fig:result35}
\end{figure}

\begin{figure}
    \centering
    \includegraphics[width=0.8\textwidth]{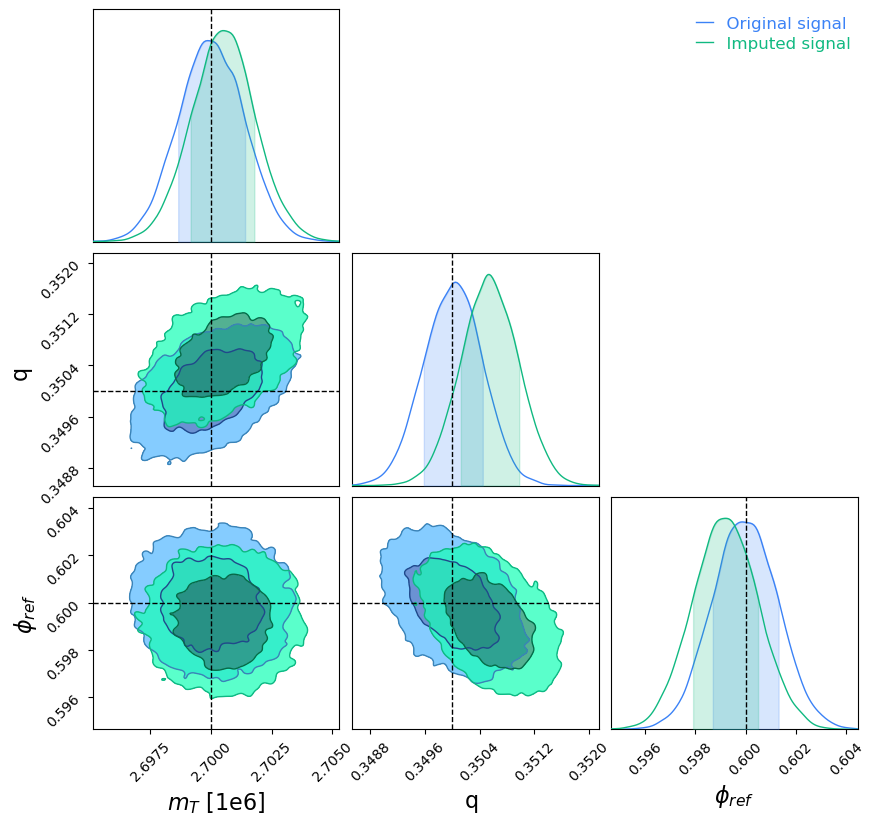}
    \caption{Parameter estimation example in MBH case for $7$-hour scheduled gaps in merging time. The black dashed line represents the true value. The shadowed areas represent the posterior incredible interval for one error in marginal distributions.}
    \label{fig:result7}
\end{figure}

\newpage
\twocolumngrid 

\bibliographystyle{unsrt}
\bibliography{refs}  

\end{document}